\documentclass[twocolumn,superscriptaddress,amsmath,amssymb,showpacs,prl]{revtex4-1}

\usepackage{graphicx}
\usepackage{color}
\renewcommand{\phi}{\varphi}

\begin{document}

\title{Theoretical prediction of a highly responsive material: spin fluctuations and superconductivity in $\text{FeNiB}_\text{2}$ system}

\author{Renhai Wang}
    \affiliation{Department of Physics, University of Science and Technology of China, Hefei 230026, China}
    \affiliation{Ames Laboratory, US DOE, Ames, Iowa 50011, USA}
\author{Yang Sun}
	\email[Email: ]{yangsun@ameslab.gov}
	\affiliation{Ames Laboratory, US DOE, Ames, Iowa 50011, USA}
\author{Vladimir Antropov}
	\email[Email: ]{antropov@ameslab.gov}
	\affiliation{Ames Laboratory, US DOE, Ames, Iowa 50011, USA}
\author{Zijing Lin}
    \affiliation{Department of Physics, University of Science and Technology of China, Hefei 230026, China}	
\author{Cai-Zhuang Wang}
	\affiliation{Ames Laboratory, US DOE, Ames, Iowa 50011, USA}
	\affiliation{Department of Physics, Iowa State University, Iowa 50011, USA}
\author{Kai-Ming Ho}
    \affiliation{Ames Laboratory, US DOE, Ames, Iowa 50011, USA}
    \affiliation{Department of Physics, Iowa State University, Iowa 50011, USA}
    \affiliation{International Center for Quantum Design of Functional Materials (ICQD), Hefei National Laboratory for Physics Sciences at the Microscale, University of Science and Technology of China, Hefei 230026, China}
	
\date{Oct. 20, 2019}

\begin{abstract}

By analyzing Fe-Ni-B compositional diagram we predict an energetically and dynamically stable $\text{FeNiB}_\text{2}$ compound. This system belongs to the class of highly responsive state of material, as it is very sensitive to the external perturbations. This state is also characterized by a high level of spin fluctuations which strongly influence possible magnetic long- and short-range orders. Furthermore, we demonstrate that these antiferromagnetically dominating fluctuations could lead to the appearance of spin mediated superconductivity. The obtained results suggest a promising avenue for the search of strong spin fluctuation systems and related superconductors.

\end{abstract}

\maketitle

\textit{Introduction -} Spin fluctuations (SF) in itinerant electron magnets play an important role in many metallic systems including weak itinerant magnets, heavy fermion compounds, actinides, Invar alloys and many magnetoresistive materials. SF also have been identified in high-temperature superconductors \cite{S1} and recently discovered Fe-based superconductors \cite{S2} with a clear suggestion about their crucial influence on the mechanisms of unconventional superconductivity. In the majority of cases superconductivity appears near magnetic quantum critical point, where magnetism has a pure itinerant character with no local moments involved. This is exactly a case when quantum SF (including spin zero-point motion) are strong and provide a dominating contribution to many observable physical properties. However, as a typical example, in a case of superconductivity, the direct calculations and prediction of critical temperature from first principle is still not possible. In this situation one can focus on searching for new materials with a strong level of SF, leading to possible development of promising physical properties. 

\begin{figure}
\includegraphics[width=0.35\textwidth]{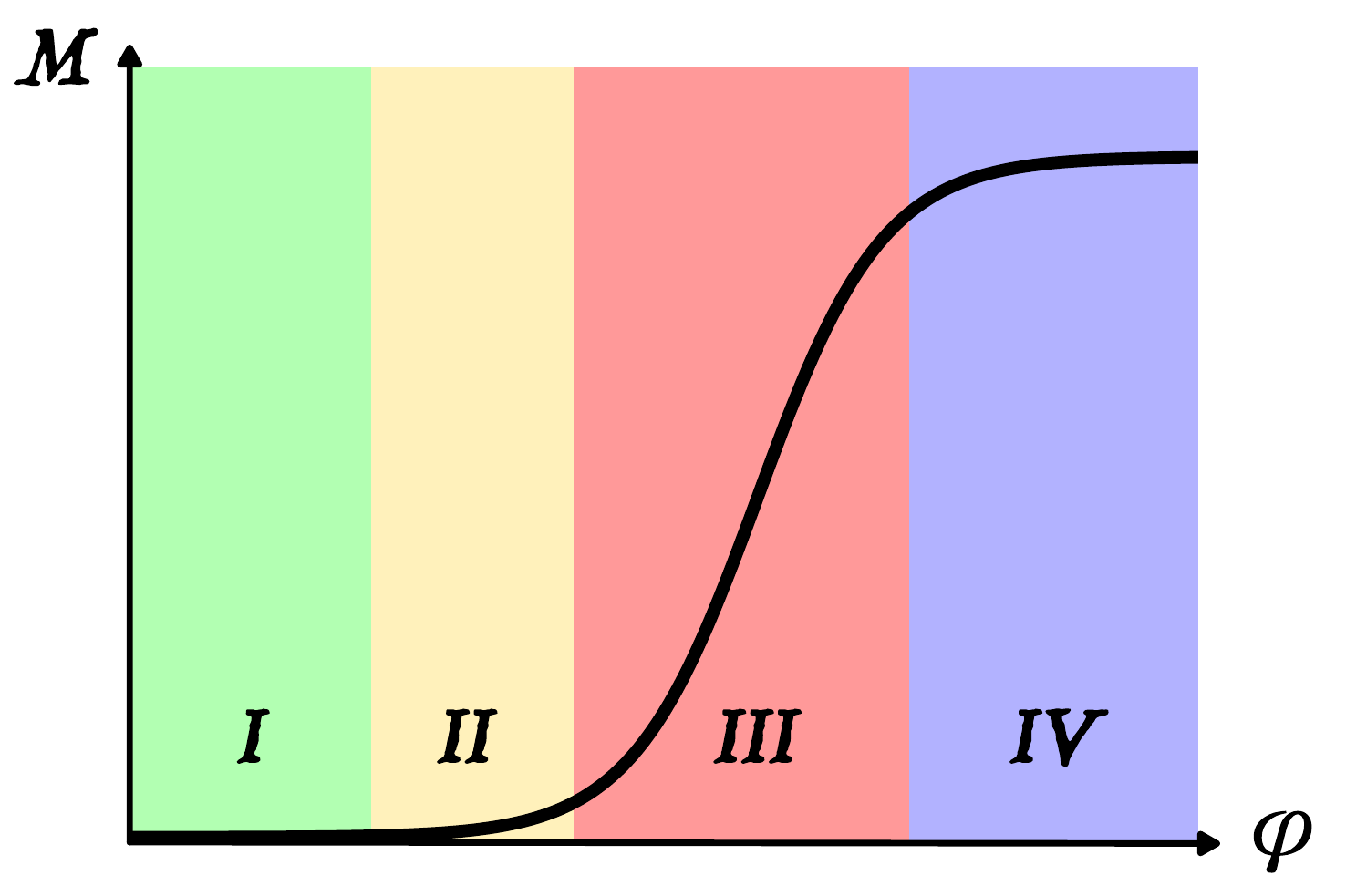}
\caption{\label{fig:fig1} Schematics of HRS from a viewpoint of mean field description. Regions I and IV represent nonmagnetic and saturated regimes; Region II corresponds to quantum critical point and a system in the area III is in HRS regime with possible large gradients of magnetization and strong spin fluctuations. $M$ is the order parameter while $\phi$ is the external perturbation (pressure, magnetic field, etc).}
\end{figure}

Here we will introduce qualitatively the concept of a highly responsive state (HRS) (see Fig.~\ref{fig:fig1}). This term is used here to indicate that a physical system is in a state characterized by an order parameter which is very sensitive to the external perturbations (temperature, pressure, magnetic fields, etc.). The HRS state is naturally related with a closeness to the instability and, as usual in such situations, one can expect that fluctuations are very important. 

To search for new material of such type we start our study with a less known system Fe-Ni-B in the concentration regime where magnetism is not supposed to be very strong (equal concentration of magnetic 3d atoms and nonmagnetic B atoms). Further we proceed with a structural search for new stable compounds and then study the stability of predicted antiferromagnetic (AFM) order by adding possible SF in the ground state (quantum zero-point motion). At the end we estimate the influence of quantum SF on possible appearance of SF induced superconductivity. 

\textit{Structure prediction -} Crystal structures of Fe-Ni-B were searched by an adaptive genetic algorithm (AGA), which integrates auxiliary interatomic potentials and first-principles calculations together in an adaptive manner to ensure the high efficiency and accuracy (see Supplementary Materials for more details) \cite{S3,S4}. The structure searches were only constrained by the chemical composition, without any assumption on the Bravais lattice type, symmetry, atom basis or unit cell dimensions. A range of different composition around the FeNiB (111) compositional ratio (i.e., 211, 121, 112, 311, 131, 113, 221, 122, 212, 133, 313, 331, 321, 312, 123, 132, 231, 213, 223, 232, 322, 233, 323 and 332) were selected with 2 and 4 formula units to perform the AGA search. First-principles calculations were carried out using the projector augmented wave (PAW) method \cite{S5} within density functional theory (DFT) as implemented in VASP code \cite{S6,S7}. The exchange and correlation energy is treated within the spin-polarized generalized gradient approximation (GGA) and parameterized by Perdew-Burke-Ernzerhof formula (PBE) \cite{S8}. A plane-wave basis was used with a kinetic energy cutoff of $520\text{ eV}$. During the AGA search, Monkhorst-Pack's sampling scheme \cite{S9} was adopted for Brillouin zone sampling with a k-point grid of $2\pi \times 0.033 \text{ \AA} ^{-1}$, and the ionic relaxations stopped when the forces on every atom became smaller than $0.01 \text{ eV}/\text{\AA}$. The energy convergence criterion is $10^{-4}\text{ eV}$. The phonon calculations were performed with the finite difference method via Phonopy code \cite{S10}. The magnetic properties were calculated in a higher-quality calculation using a finer k-point sampling grid of spacing $2\pi \times 0.02 \text{ \AA} ^{-1}$ and a stricter energy convergence criterion of $10^{-6}\text{ eV}$. 

\begin{figure}
\includegraphics[width=0.48\textwidth]{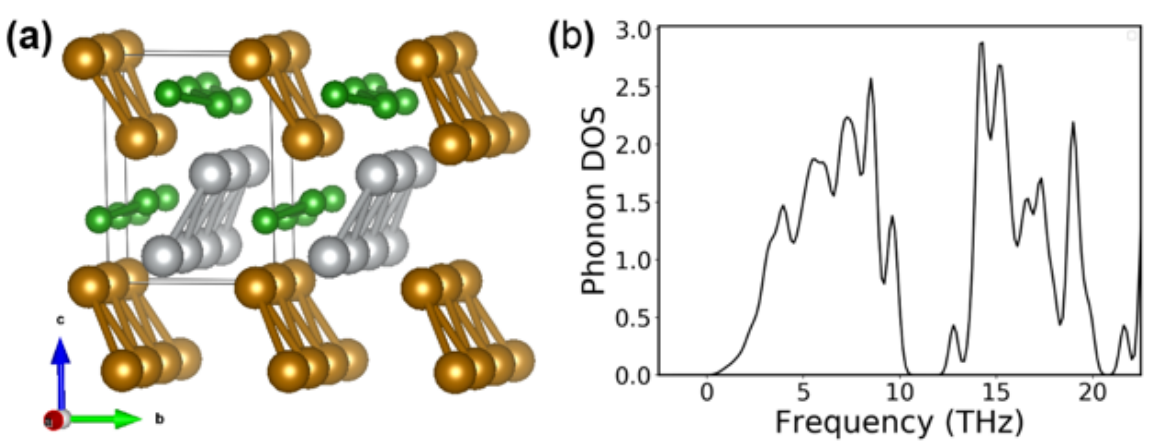}
\caption{\label{fig:fig2} (a) The configuration of $\text{FeNiB}_\text{2}$ from AGA search. Atom color: Fe atoms colored with brown; Ni atoms colored with silver; B atoms colored with green. (b) Phonon density of States.}
\end{figure}

Among all AGA searched ternary phases, a new $\text{FeNiB}_\text{2}$ phase, with 2 formula units and a space group of $P2_1/m$ (space group number 11), is found to show the energy even smaller than the Gibbs triangle formed by the existing stable phases on the ternary phase diagram. The structure of this new phase is shown in Fig. ~\ref{fig:fig2}(a). There is one Wyckoff site for Fe, Ni and two Wyckoff sites for B, respectively (the crystallographic details are given in Supplementary Materials). The Fe, Ni and B atoms formed separated 1D chain. Phonon calculation further confirmed the phase is dynamically stable without any soft phonon mode at $T=0 \text{ K}$, which is shown in Fig. ~\ref{fig:fig2}(b).

With the newly found $\text{FeNiB}_\text{2}$ structure, the existing Fe-Ni-B phase diagram should be updated. In the current DFT-calculated database (i.e., the Materials Project \cite{S11}), there are eight reported binary stable phases however there is no ternary stable phase reported. It should be noted that while the experimental database (i.e. Inorganic Crystal Structure Database \cite{S12}) shows that a few ternary phases such as FeNiB, Fe$_\text{2}$NiB, Fe$_\text{3}$Ni$_\text{3}$B$_\text{2}$ and Fe$_\text{3}$Ni$_\text{20}$B$_\text{6}$ previously synthesized by experiments, they are all metastable phases according to DFT calculations, indicating they may decompose to the surrounding stable phases on the Gibbs triangle in the equilibrium state \cite{S13,S14}. The existing ternary convex hull computed from DFT calculation are shown in Fig.~\ref{fig:fig3}(a) (the detailed energy values are provided in Supplementary Materials). At the composition of $\text{FeNiB}_\text{2}$, the thermodynamic stability is determined by the Gibbs triangle formed by the nearby Ni$_\text{4}$B$_\text{3}$, FeB, and FeB$_\text{2}$ phases. Considering the following reaction 
$ \text{FeNiB}_\text{2} \rightarrow (\text{Ni}_\text{4}\text{B}_\text{3}+3\text{FeB}+\text{FeB}_\text{2})/4$, 
$\text{FeNiB}_\text{2}$ is a stable phase if the formation energy of the left hand side is lower than that of the right hand side. The DFT calculations indeed show that the formation energy of $\text{FeNiB}_\text{2}$ in the $P2_1/m$ structure is equal to $-340.9 \text{ meV/atom}$, while at the right-hand side, i.e. the energy on the convex hull at the composition of $\text{FeNiB}_\text{2}$, is calculated to be $-326.57 \text{ meV/atom}$. Therefore, $\text{FeNiB}_\text{2}$ is thermodynamically stable at the temperature of $0 \text{ K}$ and the convex hull of the Fe-Ni-B system should be updated with the inclusion of the new $\text{FeNiB}_\text{2}$ phase. The new phase diagram of Fe-Ni-B system at $0 \text{ K}$ are show in Fig.~\ref{fig:fig3}(b). 

\begin{figure}
\includegraphics[width=0.5\textwidth]{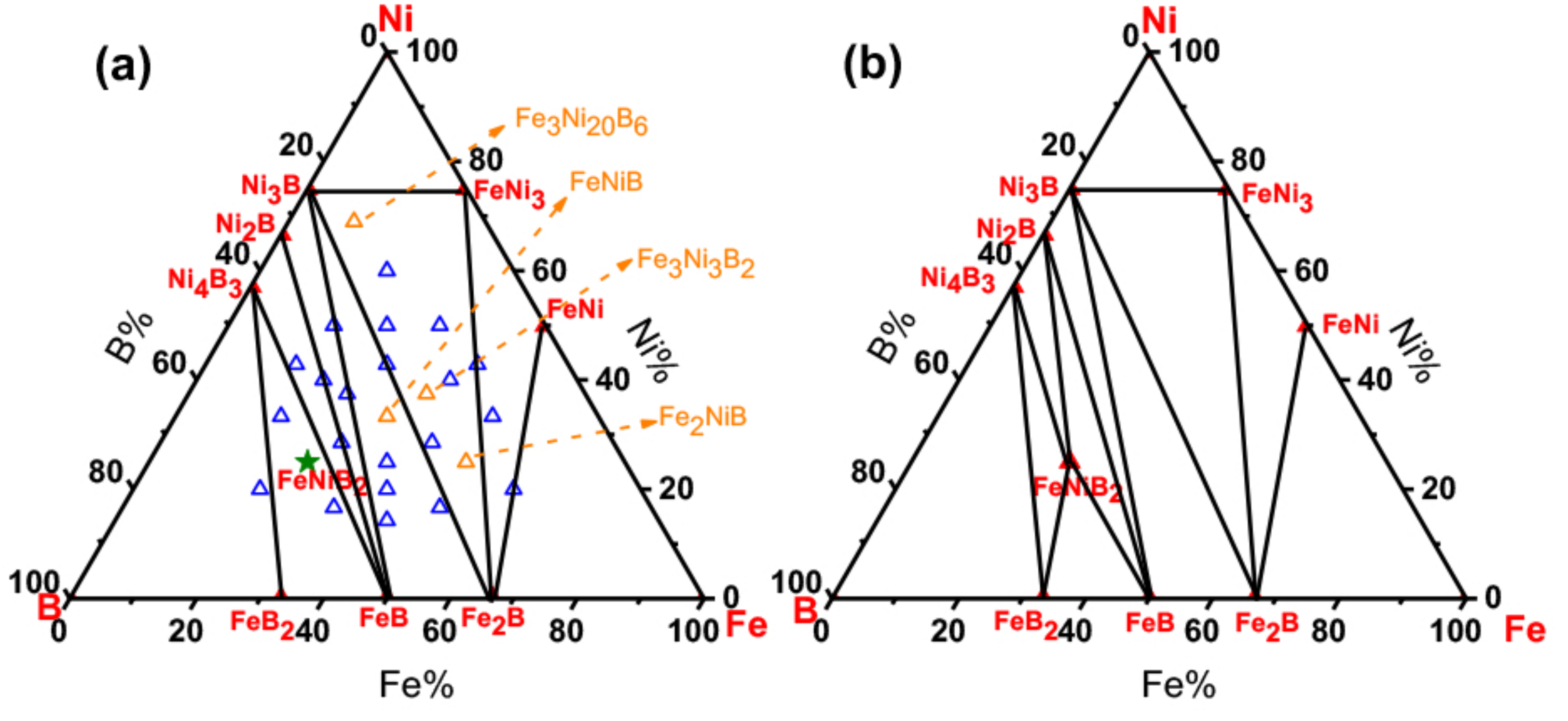}
\caption{\label{fig:fig3} DFT-calculated convex hull of the Fe-Ni-B system based on (a) previously reported phases and (b) after including the newly discovered $\text{FeNiB}_\text{2}$ phase. All DFT calculations were carried out at $0 \text{ K}$. Red solid triangles represent stable compounds and the green solid star represent newly found stable phase, while blue and orange open triangles indicate the metastable phases from AGA search and experiments, respectively. The black lines separate the compositional space to Gibbs triangles.}
\end{figure}

Considering the stability of $\text{FeNiB}_\text{2}$ phase, we search the literature to see if there is any hint of its existence. First, we find the stable CoB phase (tetragonal, space group $Pnma$) shows a same motif as the current $\text{FeNiB}_\text{2}$ phase. From the chemical viewpoint, Co atoms should easily be replaced by Fe and Ni atoms. Moreover, while the ground state of FeB is the tetragonal phase ($I4_1/amd$), experiments always result the metastable FeB analogous to the CoB structure, due to the vibrational properties as suggested in Ref. \cite{S15n}. Very interestingly, Gianoglio and Badini \cite{S14} reported that Ni can substitute Fe in the FeB-Pnma phase up to 70\%. They showed the lattice of 50\% Ni substituted FeB were 0.40, 0.30 and 0.54 nm, which is consistent with our current results of 0.395, 0.30 and 0.532 nm except that the cell angle $\beta$ becomes $91.35^{\circ}$ in our DFT calculations. With these evidences, we believe the $\text{FeNiB}_\text{2}$ exists however it is not explicitly reported by experiments. 

\begin{figure}
\includegraphics[width=0.48\textwidth]{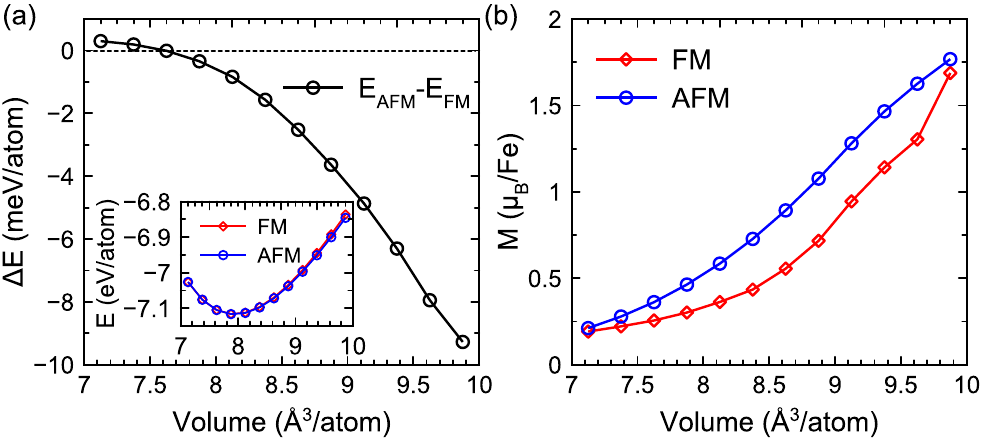}
\caption{\label{fig:fig4} (a) The energy difference and (b) the Fe magnetic moment at AFM and FM states as a function of the volume for the $\text{FeNiB}_\text{2}$ structure. The insert in (a) shows the total energies for both configurations.}
\end{figure}

With the verified stability of the new $\text{FeNiB}_\text{2}$ structure we are now in position to investigate its magnetic ground state. The DFT calculations are performed for the ferromagnetic (FM) state and the simplest AFM state, in which two Fe atoms in the primitive cell show opposite spin directions. It shows a strong competition of energetic stability between the FM and AFM states in the $\text{FeNiB}_\text{2}$ structure at $0\text{ K}$. As shown in Fig.~\ref{fig:fig4}(a), the two states exhibit very similar energies, while the AFM state has a slightly lower energy $\sim 0.45 \text{ meV/atom}$ at the equilibrium volume $7.8 \text{\AA}^3\text{/atom}$. As the volume increases, the energy difference becomes larger and the AFM state is always slightly more stable than FM state. More interestingly, in Fig.~\ref{fig:fig4}(b), the equilibrium magnetic moments demonstrate a rapid increase (nearly by 10 times). While the current results were based on a fixed geometry of the lattice, the calculation with relaxed structures shows almost same results (see Supplementary Material). By using the mean field approximation of Heisenberg model, the Neel temperature is estimated to be nearly 50-60K (the Heisenberg model parameters were calculated using Ref. \cite{S16n}). It suggests the $\text{FeNiB}_\text{2}$ structure is locating in the highly responsive state (region III in Fig.~\ref{fig:fig1}). Coexistence of different magnetic long-range orders even for small values of atomic magnetic moments indicates that energy profile should have many local minima without well-defined global minimum. Correspondingly, the system is very sensitive to any external perturbation and could easily change its magnetic state (“magnetic chameleon”) including magnetic tunneling. We note that this HRS can be a characterization of $\text{FeNiB}_\text{2}$ phase compared to the analogous FeB-Pnma phase which has a ferromagnetic ground state.

\textit{The effects of zero-point spin fluctuations on the magnetic stability of the ground state of FeNiB$_\text{2}$ -} Let us first address the issue of magnetic instability in $\text{FeNiB}_\text{2}$ . Our nonmagnetic calculations of this system revealed that the density of states (DOS) at the Fermi level is dominated by Fe electrons. So the Stoner criterion is fulfilled \begin{equation}
1-I\chi_0>0
\label{eos}.
\end{equation}
Here $\chi_0$ is non-enhanced magnetic susceptibility, while $I$ is the effective Stoner parameter of static LDA. For multicomponent system the corresponding Stoner criterion was obtained in Ref. \cite{S15} and is written as
\begin{equation}
N_t\left(E_F\right)I>1 \text{ or } \sum_{i}{I_iN_i^2}>N_t
\label{eos},
\end{equation}
where the effective Stoner parameter $I$ for multicomponent system is defined as $I=\sum_{i}{I_i\left(\frac{N_i}{N_t}\right)^2}$, where $I_i$ and $N_i$ are partial Stoner parameter and partial DOS at the Fermi level, correspondingly, and $N_t$ is the total DOS at the Fermi level. The obtained numbers are shown in Table \ref{table:tab1} and the resulting criteria of possible long-range AFM order is fulfilled. Such criteria are also fulfilled at many other $q$-vectors, including FM one. We also found that criteria of a local magnetic moment stability (Anderson criteria \cite{S17,S19n}) is satisfied as well.

\begin{table}
\caption{\label{table:tab1} The partial density of state at the Fermi level and the Stoner parameter at each atom for $\text{FeNiB}_\text{2}$.}
\centering
\begin{tabular} {  c | c | c   }
\hline
\hline
\textbf{ } & DOS @ E$_{f}$ (State/eV) & $I$ (eV) \\
\hline
Total ($N_t$) & 1.962 &  \\
\hline
Fe & 1.477 & 0.84  \\
\hline
Ni & 0.312 & 0.87  \\
\hline
B & 0.0895 & 2.20  \\
\hline
\hline
\end{tabular}

\end{table}

However, LDA traditionally overestimates the magnetism for weakly magnetic systems due to neglect of quantum SF at $T=0\text{ K}$ (spin zero-point motion). To study the stability of predicted above weak magnetism in $\text{FeNiB}_\text{2}$ we used a modified version of SF theory from Ref.\cite{S16}. First, we obtain the $ab ~ initio$ SF energy (absent in LDA) using RPA technique \cite{S17}. Then we calculate a corresponding SF contribution to the spin susceptibility and finally obtain renormalized Stoner criterion for magnetic instability. 

The following RPA expression for the SF contribution to the total energy has been used 

\begin{widetext}

\onecolumngrid
\begin{equation}
E_{SF}=\lim_{T\rightarrow0} F_{SF}=2\hbar\lim_{T\rightarrow0}{\sum_{v}\int_{0}^{I}{dI\int_{\Omega_{BZ}}{d\pmb{q}\int_{-\infty}^{\infty}d\omega\left(\mathrm{Im}\chi_v\left(\pmb{q},\omega\right)-\mathrm{Im} \chi_{v0}\left(\pmb{q},\omega\right)\right)\mathrm{coth} \left(\beta\omega/2\right)}}}
\label{eos}.
\end{equation}

\end{widetext}

\twocolumngrid

This expression can also be rewritten as 
\begin{equation}
E_{SF}=\frac{1}{2}\int_{0}^{I}dI\left(\left\langle M^2\right\rangle-\left\langle M_0^2\right\rangle\right)
\label{eos},
\end{equation}
where $\left\langle M^2\right\rangle$ is a square mean of magnetic moment obtained with enhanced and nonenhanced susceptibility, correspondingly (see Fig.8 in Ref.\cite{S18} for detailed definition and calculations of these quantities in pure 3d metals). Computational details of our linear response method can be found in Ref.\cite{S18}.

The transverse part of SF correction to the inverse magnetic susceptibility is obtained directly as  
\begin{equation}
\chi_{SF}^{-1}=\frac{1}{2M}\frac{\partial}{\partial M}\left(\int_{0}^{I}dI\left(\left\langle M^2\right\rangle-\left\langle M_0^2\right\rangle\right)\right)
\label{eos}.
\end{equation}

This expression is different from the one used in Refs.\cite{S16,S19} as we do not employ low frequency and long wavelength approximations \cite{S20}. The SF renormalized Stoner criterion then is written as
\begin{equation}
\left(\chi_{SF}^{-1}+\chi^{-1}\right)\chi_0=1-I^\ast\chi_0>0
\label{eos},
\end{equation}
where $I^\ast$ is a renormalized Stoner parameter, which includes quantum spin zero points SF beyond usual static local density approximation. This renormalization was first noticed by Moriya  \cite{S16} and studied recently using electronic structure calculations in Ref. \cite{S19} for several simple 3d metals. All these studies have been performed using \textit{ad-hoc} low frequency and long wave approximations. Our approach is truly ab-initio and takes into account SF at all $q$ vectors and frequencies.

Our results indicated that the SF renormalization of the effective Stoner parameter in FeNiB$_\text{2}$ is nearly $7\%$ with dominating suppression coming from Fe sites. Many different SF with q-vectors around AFM instability contribute to this suppression in the frequency range up to $0.3$ eV, but overall transitions inside electronic d-band ($4-5$ eV) contribute to this renormalization. It suggests that the existence of predicted SF can be measured by neutron scattering experiments. Both longitudinal and transversal SF on Ni sites are also strong but due to small original static moment on Ni atoms, the Ni contribution to the total magnetization suppression is small. 

The appearance of AFM like magnetic instability can also be favorable for SF induced superconductivity. While s-wave and triplet superconductivity could theoretically appear near FM instability, such systems are not really known. Superconductivity reports in weakly FM system ZnZr$_\text{2}$ under pressure have not been confirmed. But more popular singlet superconductivity can naturally appear near AFM instability (cuprates \cite{S1}, iron pnictides \cite{S2}). So, the appearance of superconductivity looks possible in this system and now we will estimate a strength of electron-spin interaction and its contribution to SF induced superconductivity.

\textit{The effects of zero-point spin fluctuations on superconductivity in FeNiB$_\text{2}$ -} From practical point of view, e.g., for the search of new superconducting material, the qualitative estimation of $\lambda$ is of primary interest. To obtain the required strength of spin-electron interaction we will use well known s-d exchange model \cite{S21}. To estimate $\lambda$ we do not need to consider the superconducting phase since it enters just the mass renormalization in the normal state and one needs just to deal with the corresponding expression for the electron self-energy \cite{S19},

\begin{widetext}

\begin{equation}
\Sigma_\mathrm{k}\left(E\right)=I^2\sum_{q}{\int_{0}^{\infty}\frac{d\omega}{4\pi}Im\chi\left(\pmb{k}-\pmb{q},\omega\right)\left[\frac{1-f_{q}}{E-\varepsilon_{q}-\omega}+\frac{f_{q}}{E-\varepsilon_{q}+\omega}\right]}
\label{eos},
\end{equation}
\end{widetext}
where $f_{q}$ is the Fermi distribution with $\varepsilon_{q}$ being one electron spectrum. The expression for the coupling constant at $T=0~\text{K}$ is then written in standard way as

\begin{equation}
\lambda=-\left.\frac{\partial\left\langle\Sigma_\mathrm{k}\left(E\right)\right\rangle}{\partial E}\right|_{E=0}
\label{eos}.
\end{equation}

To make the simplest possible estimation of electron magnon interaction one can replace the average over the Fermi surface above by the average over the Brillouin zone. To further obtain the temperature of the superconducting transition using the approach advocated above we use usual MacMillan formula \cite{S22}. Under these approximations our calculated value of $\lambda = 0.6$ and the effective SF frequency is $\omega = 0.2$ eV. With these numbers we obtain a value of $20-30$ K for the critical temperature. Thus, our qualitative estimation predicts not only suppression of magnetic long-range order in the normal state but also suggest a very promising scenario for the SF induced superconductivity near AFM quantum critical point.

In summary, we performed a structure search for a stable system near magnetic instability. Our AGA approach which integrates auxiliary interatomic potentials and first-principles calculations found a stable compound FeNiB$_\text{2}$, which appears to be a magnetically highly responsive system close to antiferromagnetic instability. The structure of FeNiB$_\text{2}$ is analogous to the FeB-Pnma phase and previous literature provide hints of its possible existence. Compared to the ferromagnetic long-range order of FeB, this system is characterized by strong anti-ferromagnetic-like spin fluctuations at $T=0$K (quantum spin zero-point motion) and possible spin fluctuation induced superconductivity. The experimental verifications of our prediction are highly desirable. Our magneto-structural combined studies thus open a promising route for the search of strong SF state and superconductors.

\begin{acknowledgments}
\textit{Acknowledgements} Work at Ames Laboratory was supported by the U.S. Department of Energy, Basic Energy Sciences, Materials Science and Engineering Division, under Contract No. DEAC02-07CH11358, including a grant of computer time at the National Energy Research Scientific Computing Center (NERSC) in Berkeley, CA. Work at University of Science and Technology of China was supported by the National Natural Science Foundation of China (11574284 \& 11774324) and the Supercomputing Center of the University of Science and Technology of China.
\end{acknowledgments}

\bibliographystyle{apsrev4-1}

\end{document}